\documentclass[aps,prl,twocolumn,superscriptaddress]{revtex4}
\usepackage{amsfonts}
\usepackage{amsmath}
\usepackage{amssymb}
\usepackage{graphicx}
\setcitestyle{super}

\begin{document}

\title{Stimuli Thresholds for Isomerization-induced Molecular Motions in Azobenzene Containing Materials.}

%\date{\today }

\author{V. Teboul}
\email{victor.teboul@univ-angers.fr}
\affiliation{ Laboratoire de Photonique d'Angers EA 4464, Universit\' e d'Angers, Physics Department,  2 Bd Lavoisier, 49045 Angers, France}


\begin{abstract}

We use large scale molecular dynamics simulations of the isomerizations of azobenzene molecules diluted inside a simple molecular material, to investigate the effect of a modification of the cis isomer shape on the induced diffusion mechanism. To this end we simulate incomplete isomerizations, modifying the amplitude of the trans to cis isomerization. We find thresholds in the evolution of the host molecules mobility with the isomerization amplitude, a result predicted by the cage-breaking mechanism hypothesis (Teboul, V.;  Saiddine, M.; Nunzi, J.M.; Accary, J.B.  \newblock \emph{J. Chem. Phys. } {\bf 2011}, {\em134}, 114517) and by the gradient pressure mechanism theory (Barrett, C.J.;  Rochon, P.L.;  Natansohn, A.L.\newblock  {\em J. Chem. Phys.} {\bf 1998}, {\em  109}, 1505-1516. ).  Above the threshold the diffusion then increases linearly with the variation of the chromophore size induced by the isomerization.

\end{abstract}

\maketitle

\includegraphics[scale=0.33]{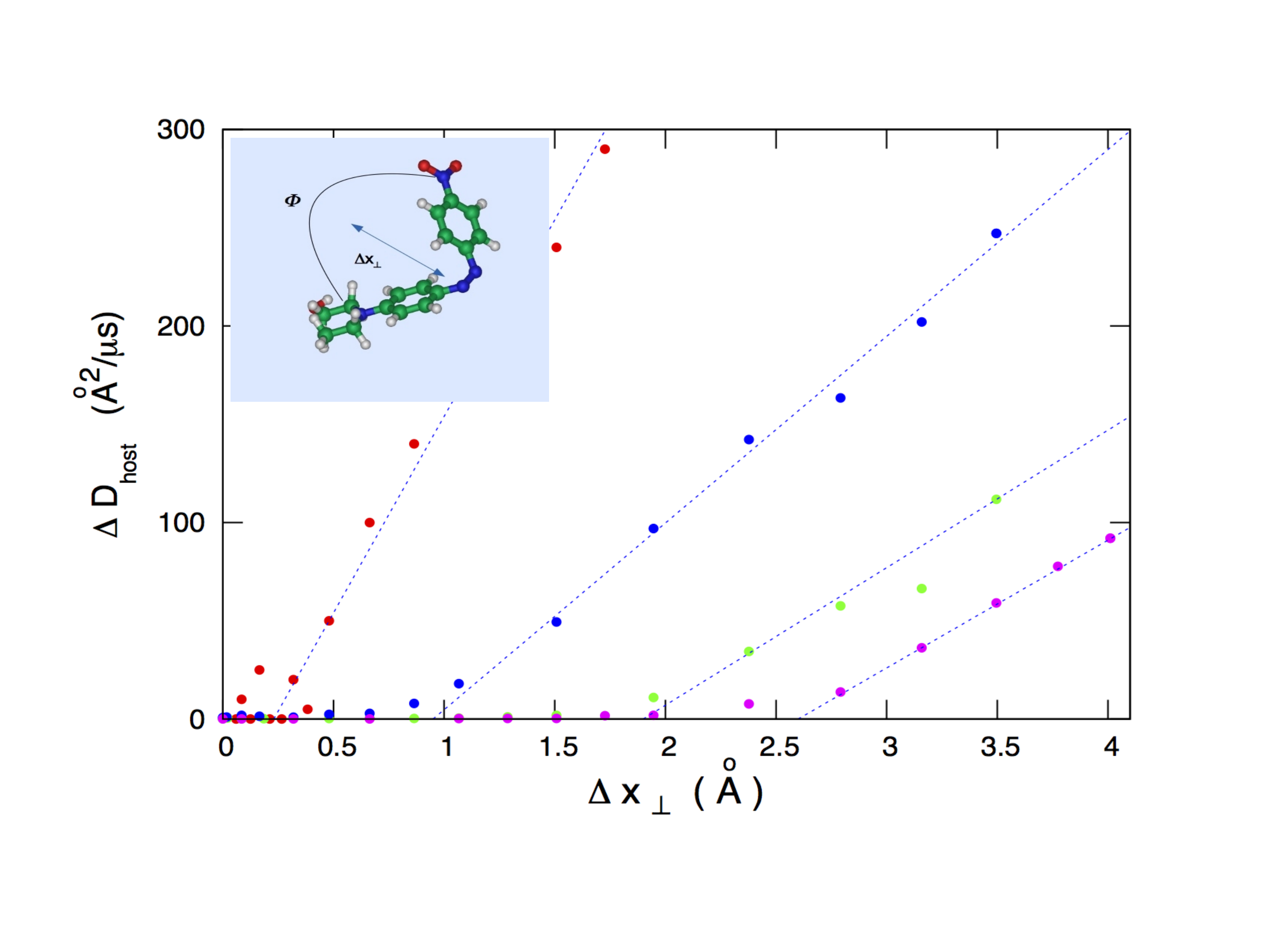}

Transport; Surface Relief Gratings;\\

\section{Introduction}

Under the nanoscale stimuli induced by the isomerization of diluted azo-dyes, soft matter undergoes intriguingly large macroscopic transport phenomena\cite{review}.
The subject has attracted a large number of investigations due to the various possible applications ranging from biological applications\cite{app1,app2,app3} to the information storage\cite{app4,app5,app6} and nanotechnology\cite{app7,app8,app9,app10,dif1,dif2,dif3,diamond}.
This unexplained isomerization induced transport is also of fundamental interest due to a possible connection\cite{prl,cage} with the glass-transition long standing problem\cite{gt1,gt2,gt3,gt4}.
If that transport property is without any doubt due to the photoisomerization of the dyes, the physical mechanisms that lead to that transport are still the object of conjectures. Various mechanisms have been proposed\cite{review} to explain that transport. The proposed mechanisms include the mean field induced by the dipolar attraction between the azo-dyes \cite{a17,a18}, the incident light electric field gradient\cite{a21}, the mechanical stress induced by the orientation of the dyes\cite{a19}, the pressure gradients induced inside the material by the isomerizations\cite{a22,a23}, an isomerization-induced cage breaking process around the azo-dyes\cite{cage}, then followed (or not) by the modification of spontaneous cooperative mechanisms in soft matter\cite{prl,gt3}, the periodic modification of the free-volume\cite{coh1,coh2} around the dye induced by the isomerization, the reptation of the dye along the polarization direction\cite{a24}.  
To complicate somehow these pictures, we expect different physical mechanisms to appear sequentially during the SRG formation\cite{review,a26}. For short time scales the chromophore's isomerization induces molecular rearrangements around the chromophore that lead to the motion of surrounding host molecules\cite{cage,a22,a23,a24,prl} and eventually to its own motion resulting in the rotation of the chromophore even at low temperatures when the thermal diffusion is small.  Then due to the preferential light absorption in the chromophore's dipole direction, the chromophores align themselves along a direction perpendicular to the electric field of the incident light\cite{review,a24} leading to the appearance of new physical mechanisms\cite{a17,a18,a19,a21}.
Recent experiments\cite{a26} show that the two sort of physical mechanisms (dipoles induced or isomerization induced) cohabit also for larger time scales, a result that one expects as long as there are still isomerizations in the medium.
In this work we are interested only in the short times physical mechanisms that are directly induced by the isomerizations. Note however  that the effect of the alignment has already been studied extensively\cite{a17,a18,a19,a21,review}.

Molecular dynamics (MD) simulations\cite{md1,md3} is an invaluable tool to unravel condensed matter physics phenomena\cite{md4,md5,md6,md7,v0} at the microscopic level. To shed some light on the physical mechanism at the origin of the induced transport, in this paper we use MD simulations to investigate the effect of a modification of the stimuli that are at the origin of the effect. 
For that purpose we 'artificially' tune in the simulations the amplitude of the chromophore isomerization (see Figures 1a and 1b), and then study the response generated by the corresponding variation of the isomerization-induced stimuli on the diffusion process. 
Our aim is to search for the existence of stimuli thresholds for the isomerization-induced diffusion.

In the cage breaking picture\cite{cage}, during the isomerization process, the chromophore pushes a few nearby host molecules out of their cages,  inducing diffusion.
 For that process to occur, the change of shape of the chromophore has to be large enough to push host molecules through their caging potential barrier into a new minimum of the mean field potential. As a result in the cage breaking picture  there is an isomerization amplitude threshold to induce diffusion. 
The other theory that predicts thresholds is the pressure gradient theory\cite{a22,a23}. In this theory the molecular motions are induced by the pressure gradient created on the medium by the change of the effective volume of the chromophore during its isomerization. The pressure gradient has thus, in that theory, to overpass the material stress threshold to permit the molecular motions.  
In addition to the theoretical interest for the explanation of the physical mechanism, the existence of thresholds implies minimal conditions on the stimuli to induce the motions.  
More generally we expect a study of the cis isomer shape dependence on the diffusion process to lead to optimization parameters for the surface relief grating (SRG) processes and the related applications\cite{app1,app2,app3,app4,app5,app6,app7,app8,app9,app10}.

\section{Calculations}

We simulate the photoisomerization of one ($N_{c}=1$) "dispersed red one" (DR1) molecule ($C_{16}H_{18}N_{4}O_{3}$, the probe) inside a matrix of $N=2688$, $800 $ or $300$  linear molecules (the host). 
We use periodic boundary conditions.
At constant density $\rho$, a decrease of the number $N$ of host molecules  corresponds to a decrease of the simulation box volume $v$, as $N_{c}$ is constant. 
For our smaller $N$ value ($N=300$), we found that size effects increase slightly the viscosity of the host but the resulting modifications of the dynamics in the vicinity ($R<10$ \AA) of the chromophore  are within the error margins of our results. Consequently we find a linear increase of the induced diffusion with the chromophore's concentration.
%Decreasing $N$ also permits us to study the effect of the chromophore concentration ($c=N_{c}/v=1/v$) on our results.
A detailed description of the simulation procedure can be found in previous works\cite{ivt3,ivt4}. The main difference is that, taking into account the universality of the effect\cite{review}, in this paper we simplify the host molecules as much as possible to better understand the effects of the isomerization on the medium. 
We found that simulations using Methyl methacrylate host molecules\cite{mma} as in ref. \cite{ivt3,ivt4} lead to similar threshold effects than the ones we report in this study, but we simplified the host to reach larger timescales and decrease the uncertainties of our results.
We model the host molecules as constituted of two atoms ($i=1, 2$) that do interact with the following Lennard-Jones potentials: 
$V_{ij}=4\epsilon_{ij}((\sigma_{ij}/r)^{12} -(\sigma_{ij}/r)^{6})$ with the parameters: $\epsilon_{11}= \epsilon_{12}=0.5 KJ/mol$, $\epsilon_{22}= 0.4 KJ/mol$, $\sigma_{11}= \sigma_{12}=3.45$\AA\ and $\sigma_{22}=3.28$\AA$ $.
We use the mass of Argon for each atom of the linear host molecule that we rigidly bonded fixing the interatomic distance to $d=1.73 $\AA$ $.  
%The molar mass of the molecule $A=80 g$ is   
With these parameters the host (alone or mixed with the probe) does not crystallize even during long simulation runs.
For the probe molecule, we use the same interaction potentials\cite{pot1} as in previous works\cite{cage,ivt4}.
Due to the large mass that we use for the atoms of the model molecules, our density is relatively large.
The density is set constant at $\rho=2.24 g/cm^{3}$. 
With these parameters, below $T=38 K$ the system falls out of equilibrium in our simulations, i.e. $T=38$ K is the smallest temperature for which we can equilibrate the system when the chromophore does not isomerize. 
As a result above that temperature the medium behaves as a viscous supercooled liquid in our simulations and below that temperature it behaves as a solid (as $t_{simulation}<\tau_{\alpha}$).
We evaluate the glass transition temperature $T_{g}$ to be slightly smaller  $T_{g} \approx 28 K$, from the change of the slope of the potential energy evolution with the temperature.  
We use the Gear algorithm with the quaternion method\cite{md1} to solve the equations of motions with a time step $\Delta t=10^{-15} s$. 
The temperature is controlled using a Berendsen thermostat\cite{berendsen}. We model the isomerization as a uniform closing and opening of the probe molecule shape\cite{prl,cage,ivt3,ivt4} during a characteristic time $t_{0}=1 ps$. The period of the isomerization $cis-trans$ and then $trans-cis$ is also fixed in the study $\tau_{p}=1 ns$.
In a previous work\cite{ivt4} we studied in details the effects of the isomerization period $\tau_{p}$ on our results.
We found that with this large period an isomerization doesn't influence the behavior of the system long enough to affect the next isomerization effect. That means that our period is large enough for each isomerization to be an independent process.
Due to this result we do not find any aging in our system, and we find that the diffusion is proportional to the number of isomerization per second. 
%The power of the incident light inducing the isomerization is related to the isomerization period with the simple formula:
%$P=E_{\gamma}/\tau_{p}$ with $E_{\gamma}=hC/\lambda$. For a typical experimental wavelength $\lambda=5435$ \AA\ the photon energy is $E_{\gamma}=3.7$ $10^{-19}$ $J$. The surface within which a photon can initiate the isomerization of our chromophore with this wavelength is $S\approx \pi.\lambda^{2}/4 \approx 2.3 $ $10^{7}$ \AA$^{2}$.
%With these values, from $\tau_{p}=E_{\gamma}/I.S$  we find that our period  $\tau_{p}=1ns$ corresponds to a light intensity $I\approx160 $ $mW/cm^{2}$.
%These values are however rough estimates.
During the isomerization the shape of the molecule is modified slightly at each time step using the quaternion method with constant quaternion variations, calculated to be in the final configuration after a $1ps$ isomerization. This method corresponds to opposite continuous rotations of the two parts of the molecule that are separated by the nitrogen bounding.
We model the isomerization to take place at periodic intervals whatever the surrounding local viscosity.
This approximation has recently been validated experimentally\cite{diamond} as the pressure that is necessary to stop the azobenze isomerization is very large ($P> 1$  $GPa$).
%Note that previous experiments have shown that the potential energy barrier that is necessary to stop the azobenzene isomerization is very large\cite{diamond}.

\includegraphics[scale=0.18]{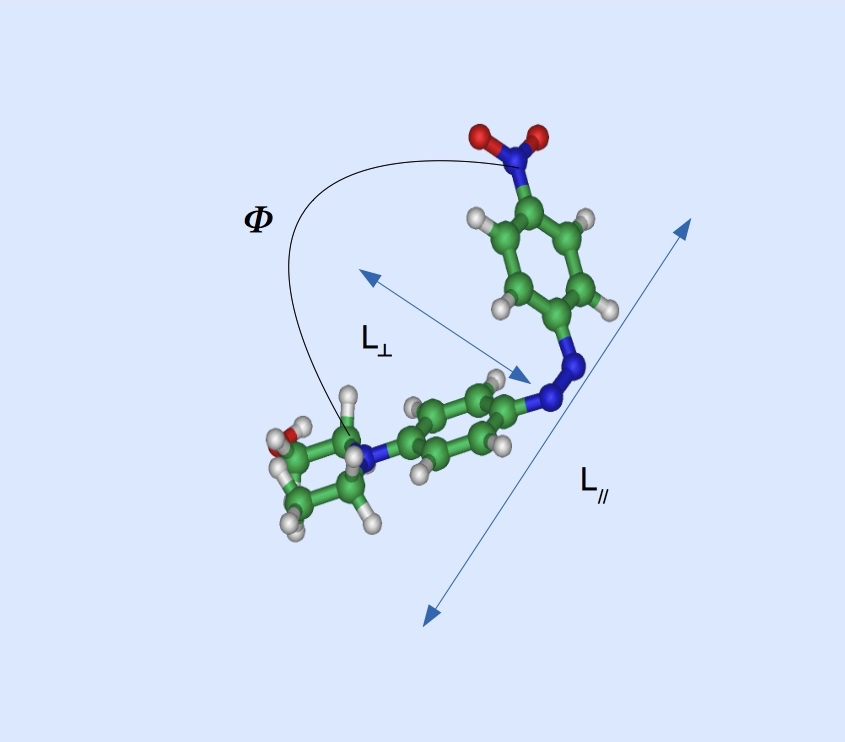}

{\em \footnotesize  FIG.1a. (color online) Opening angle $\phi$, length $L_{\parallel}$ and width $L_{\perp}$ of the chromophore. We define $\theta=\pi-\phi$ and $p=\theta/\theta_{0}$ where $\theta_{0}$ corresponds to the true $cis$ $DR1$ isomer.  The picture shows the particular cis isomer ($p=100\%$, $\theta=\theta_{0}$, $\phi=\phi_{0}$) but the resulting definitions of  $ \theta$, $L_{\parallel}$, and $L_{\perp}$ are intended to be applied for any configuration (i.e. any $p$ value). From these lengths we define $\Delta x_{\parallel}=L_{\parallel}^{trans}-L_{\parallel}$ and $\Delta x_{\perp}=L_{\perp}-L_{\perp}^{trans}\approx L_{\perp}$.}

\includegraphics[scale=0.3]{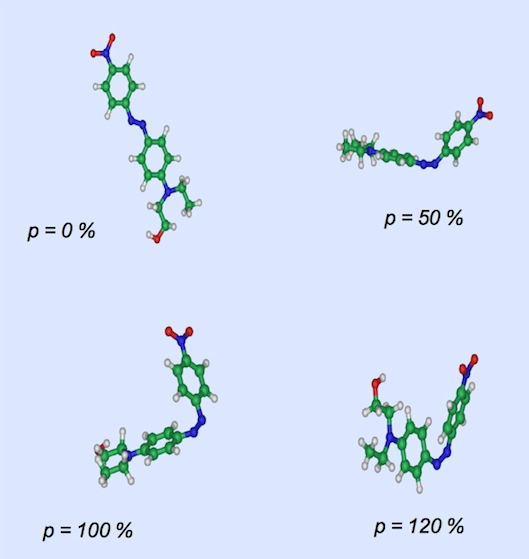}

{\em \footnotesize FIG.1b. (color online) $DR1$ chromophore molecule displayed with various amplitudes of isomerizations. 
We define the angle $\theta=\pi-\phi$ where $\phi$ is the opening angle of the $DR1$ isomer in the cis configuration.
$p=\theta/\theta_{0}$ where $\theta_{0}$ is the angle of the true $cis$ $DR1$ isomer.}

Figure 1a shows the chromophore in the cis configuration. We call $L_{\parallel}$ the length of the chromophore and $L_{\perp}$ its width as shown in the Figure, and we call  $\phi$ the opening angle of the chromophore.
Then from these quantities we define the angle $\theta=\pi-\phi$. We add the indice $0$ for quantities corresponding to the real cis isomer (i.e. to the full isomerization).
Thus $\theta=0$ when there is no isomerization and $\theta=\theta_{0}$ when the chromophore isomerizes to its real cis from.
We then define the degree of isomerization as $p=\theta/\theta_{0}$. The chromophore size modification in the direction of its axis  is $\Delta x_{\parallel}=L_{\parallel}^{trans}-L_{\parallel}$ and perpendicularly to that axis $\Delta x_{\perp}=L_{\perp}-L_{\perp}^{trans}$.

\section{Results and discussion}

\includegraphics[scale=0.33]{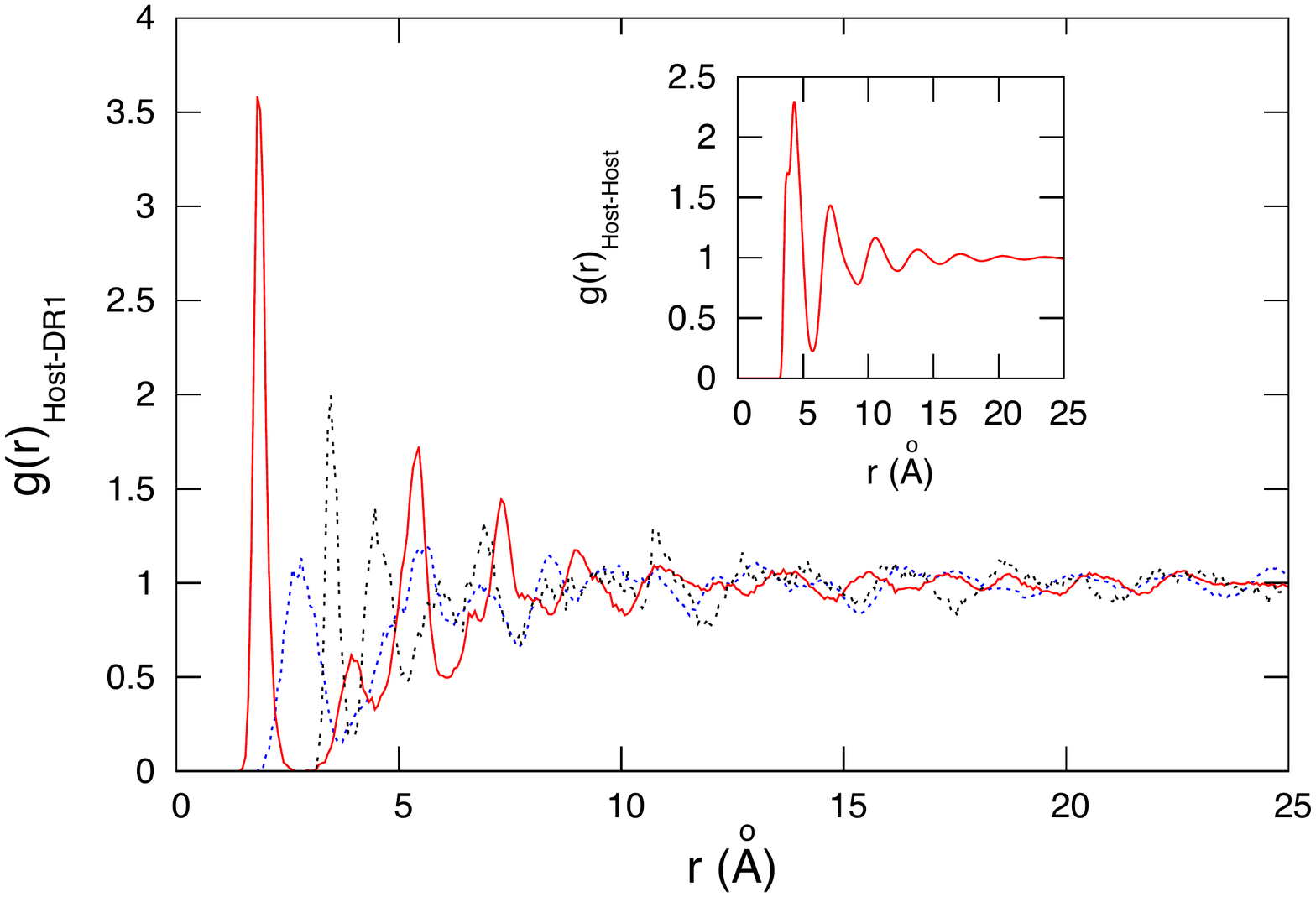}

{\em \footnotesize FIG.2a. (color online) Radial distribution function between the chromophore (cis isomer) center of mass and the hosts molecules centers of mass for various angles $\theta$ (cis isomer) i.e. various amplitudes of isomerizations $p=\theta/\theta_{0}$. 
From the left to the right hand side (first peak); Red continuous line: $p=100\%$ ($\theta=\theta_{0}$); Blue dotted line: $p=50\%$; Black dashed line: $p=0$ ($\theta=0$).
The temperature is T=40K.
Inset: Radial distribution function between the host molecules centers of mass at the same temperature.}

Figure 2a shows the radial distribution functions (RDFs) $g_{DR1-host}(r)$ between the chromophore in the cis configuration and the surrounding host molecules for various isomerization amplitudes $p$.
We see that the first peak of the RDF is shifted from $4.0$\AA\ for no isomerization (black line, $\theta=0$, $p=0$) to $1.8$\AA\  for a full isomerization (red continuous line, $\theta=\theta_{0}$, $p=1$), resulting in a $2.2$ \AA\ mean molecular displacement. The blue line in between these two peaks correspond to a $p=\theta/\theta_{0}=50 \%$ opening of the cis isomer. 
Note however that these curves represent averages taking into account the whole distribution of molecules around the $DR1$ chromophore, after the stabilization of the molecule positions, thus resulting in smaller variations of the distances than the ones transiently experienced by the most affected molecules.
The inset shows the radial distribution function $g_{host-host}(r)$ between the host molecules center of masses, and thus represents the medium mean structure. From that curve we see that a $\Delta r=2.7$ \AA\ motion (the distance between the two first peaks) is enough to push completely a molecule from the first neighbor shell to the second neighbor shell.   %that it supposes extreme pressures that are not encountered in usual materials.
%Consequently the isomerization possesses the energy required to overpass the host potential energy barriers.
The motion of a molecule to the second neighbor shell will then destabilize that shell inducing diffusion.
The radial distribution function between the dye and the host in Figure 2a shows that behavior as during the isomerization the different shells of neighbors are modified in the Figure. 

For smaller $\Delta r$ values even when not large enough to induce diffusion by its own, the induced motion may be able to help Brownian motion  to induce diffusion. That process will depend on the temperature. 

\includegraphics[scale=0.33]{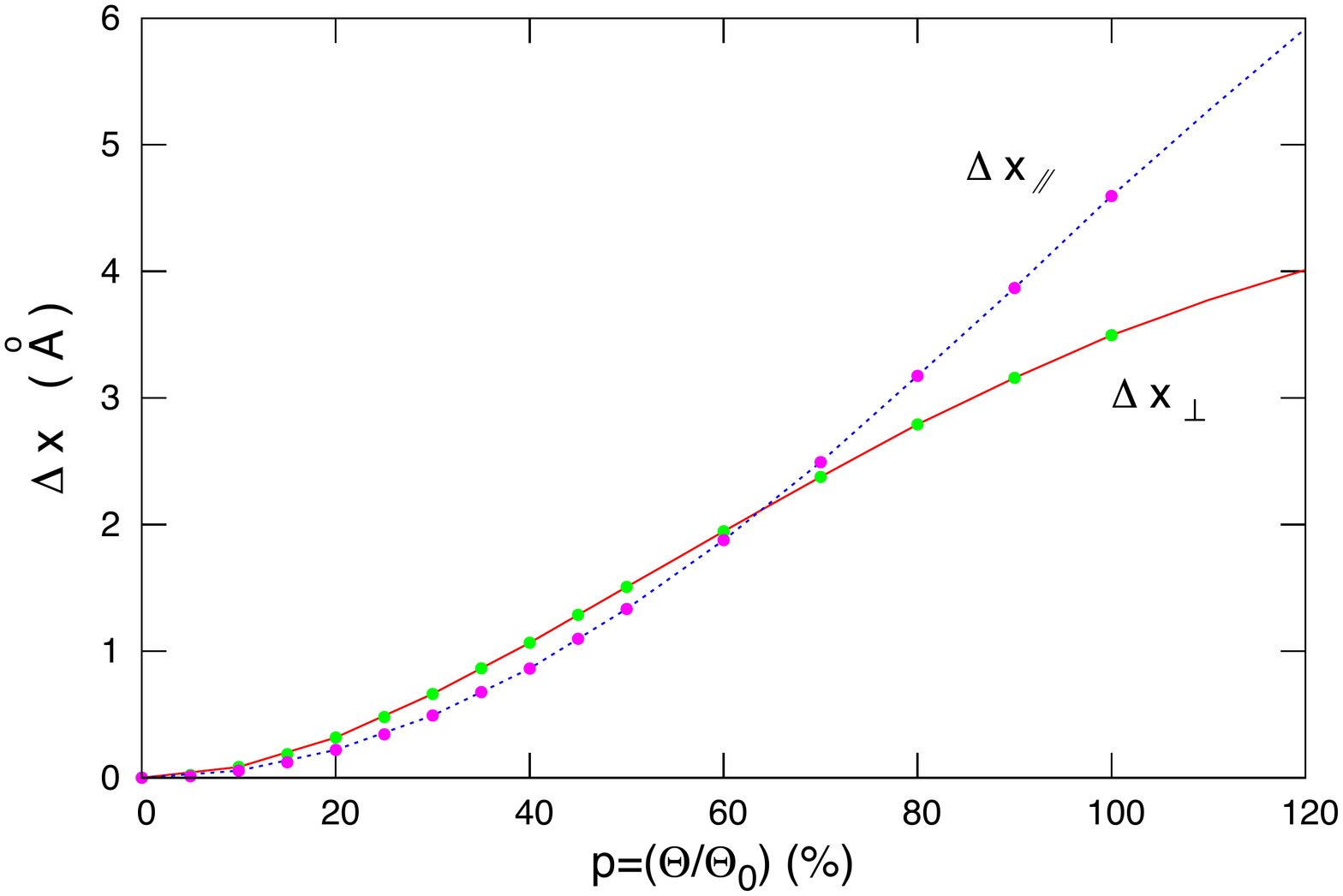}

{\em \footnotesize FIG.2b. (color online)   Evolution of the $DR1$ chromophore width  ($\Delta x_{\perp}$) and length  ($\Delta x_{\parallel}$) with the angle $\theta$ (i.e. the isomerization amplitude $p$).    $\theta_{0}$ is the angle of the true $DR1$ cis isomer. Note that as the motion is not planar the angle $\theta$ is not sufficient to calculate $\Delta x_{\perp}$ and $\Delta x_{\parallel}$, however as we use the same closing process in all the simulations, each value of $\theta$ corresponds in our work to only one value of  $\Delta x_{\perp}$ or $\Delta x_{\parallel}$.}

To evaluate the transient motions that generates the $DR1$ isomerizations inside the medium, we show in Figure 2b the evolution of the chromophore width ($\Delta x_{\perp}$) and length  ($\Delta x_{\parallel}$) with the characteristic angle $\theta$ (displayed as percentages of  the characteristic angle $\theta_{0}$ of the true $cis$ isomer of the $DR1$ molecule). We obtain these values from the locations of the different atoms of the "partial cis" chromophore molecule.
The Figure shows that the distances $\Delta x_{\perp}$ and $\Delta x_{\parallel}$ evolve more slowly than the characteristic angle $\theta$ for small values of $\theta$. Displaying the mobilities versus $\theta$ will thus amplify artificially the thresholds. To be more accurate in the following we will thus systematically plot our curves as a function of the change in size of the chromophore ($\Delta x_{\perp}$ and $\Delta x_{\parallel}$) instead of the characteristic angle. 
The Figure shows that $\Delta x_{\perp}$ varies between $0$ and $4$ \AA\ and $\Delta x_{\parallel}$ between $0$ and  $6$ \AA\ .
There is thus no doubt that, for large enough  $\theta$ values,  the $DR1$ chromophore pushes some host molecules at distances larger than the position of the potential barrier of the cage.   As a result an isomerization-induced cage breaking mechanism is clearly possible in the conditions of our study.

\includegraphics[scale=0.33]{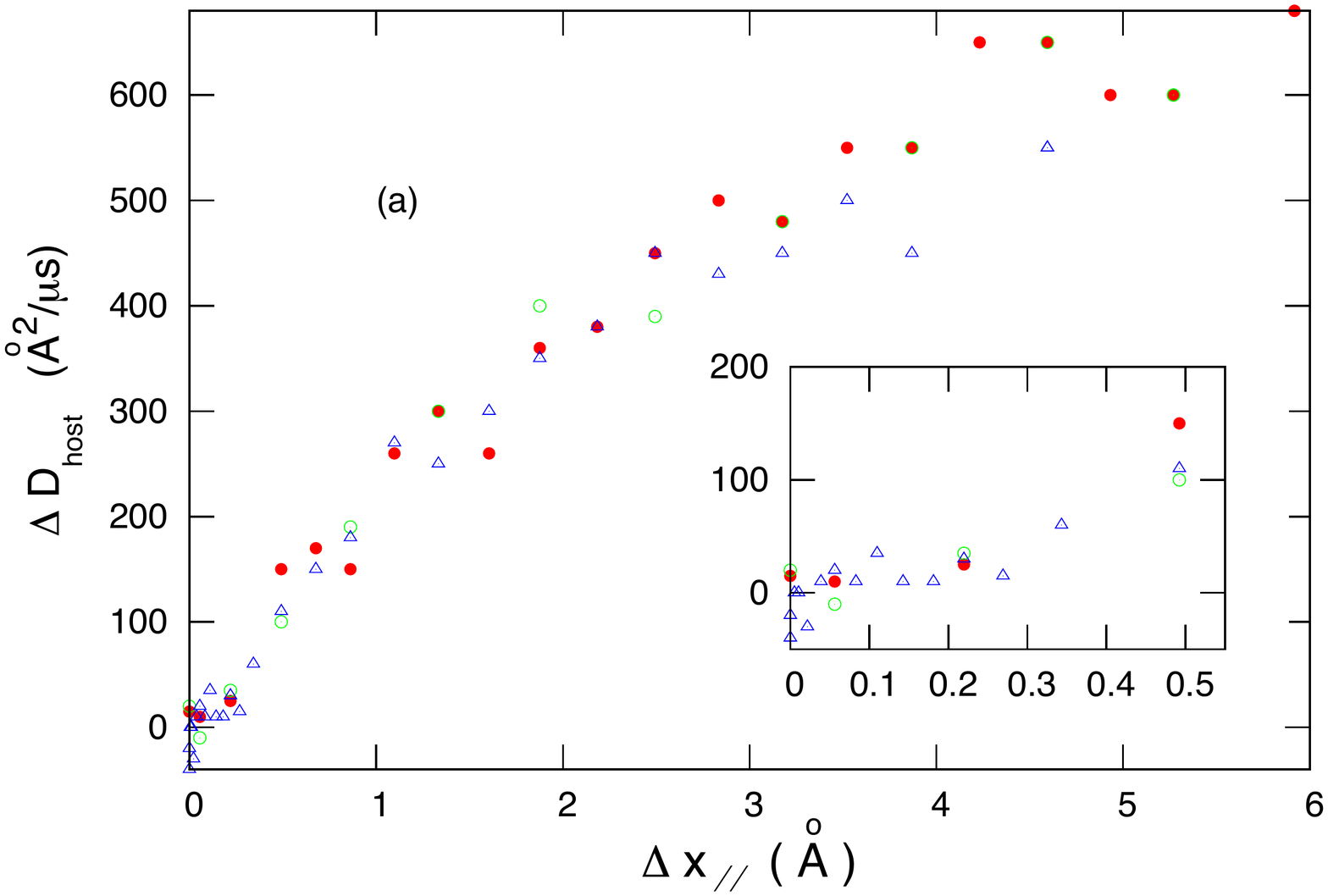}

{\em \footnotesize FIG.3a. (color online) Diffusion coefficient for host molecules, obtained from various isomerization amplitudes $p$ and displayed as a function of $\Delta x_{\parallel}$. The different symbols correspond to different simulations.
Red full circles ($N=300, t _{run} = 200 ns$); Blue empty triangles and green empty circles ($N=800, t _{run} = 100 ns$) the calculation of $D$ is restricted to host molecules located at a distance $R<10$\AA$ $ from the chromophore. The temperature in our model is T=40 K ($T>T_{g}$).
Inset: Details of the same curves for small $\Delta x_{\parallel}$ values, showing the existence of a threshold.}

Figures 3 show for various temperatures the evolution of the host molecules diffusion  with the isomerization amplitude $p$.
The Figures show the existence of a threshold displacement value $\Delta x_{\perp}^{threshold}$, a result in good agreement with the cage breaking picture\cite{cage} and the gradient pressure theory\cite{a22,a23}. 
For larger  isomerization amplitudes (i.e. induced displacements $\Delta x_{\perp}>\Delta x_{\perp}^{threshold}$) the diffusive motions increase linearly with $\Delta x_{\perp}$ while it follows a slightly more complicate evolution with $\Delta x_{\parallel}$. 

\includegraphics[scale=0.33]{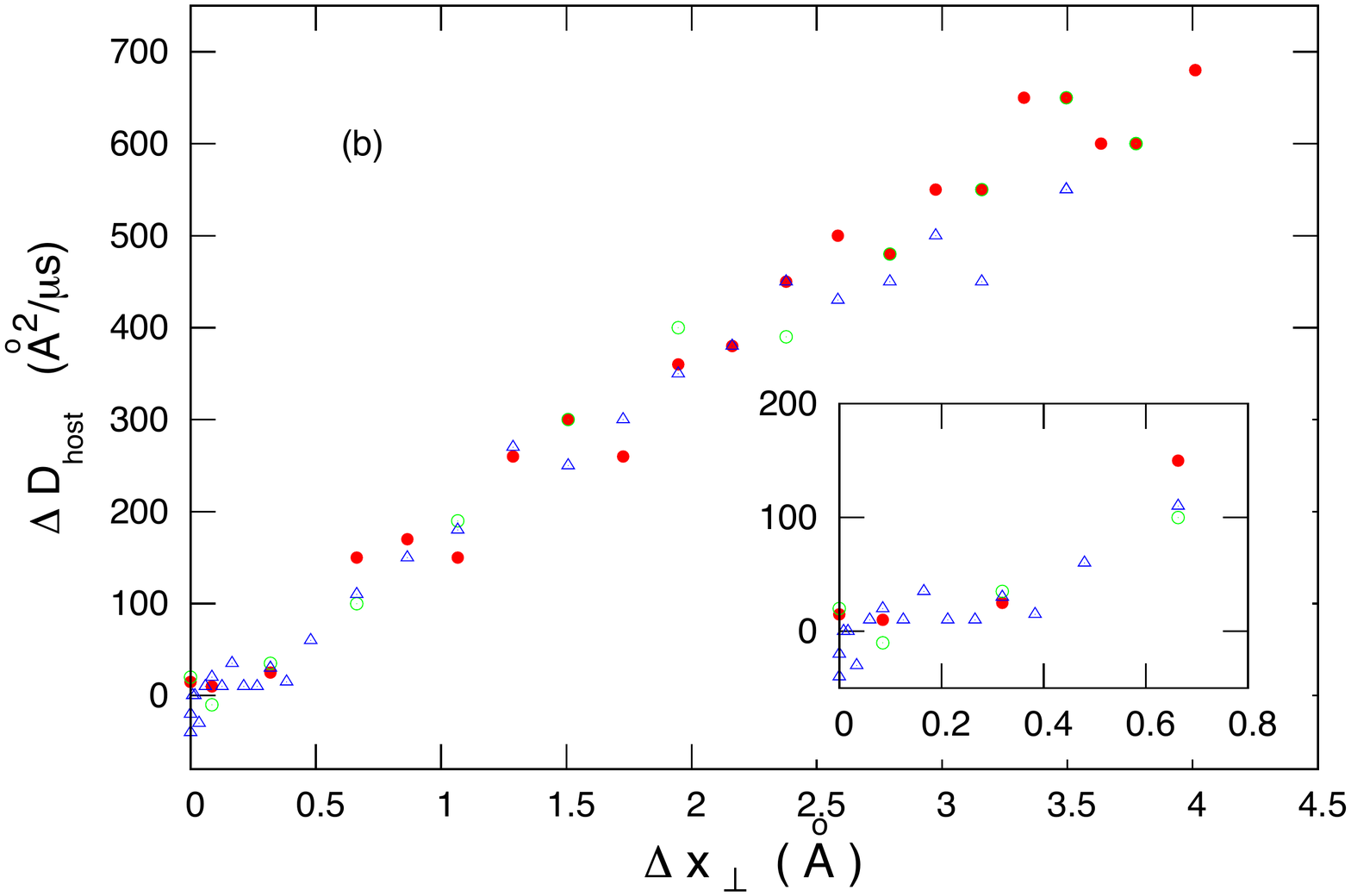}

{\em \footnotesize FIG.3b. (color online) As in Figure 3a but the diffusion coefficient is here plotted versus the chromophore width $\Delta x_{\perp}$.
Inset: Details of the same curves for small $\Delta x_{\perp}$ values, showing the existence of a threshold.\\}

This comportment is particularly visible from a comparison of Figures 3a and 3b above $T_{g}$ however we found it less apparent at lower temperatures. Nevertheless these differences between the evolutions  suggest that the perpendicular motion of the chromophore (measured by $\Delta x_{\perp}$)  mainly induces the diffusion, because  the linear dependence of $\Delta D$ with $\Delta x_{\perp}$ suggests that $\Delta x_{\perp}$ is the determinant parameter. 

\includegraphics[scale=0.33]{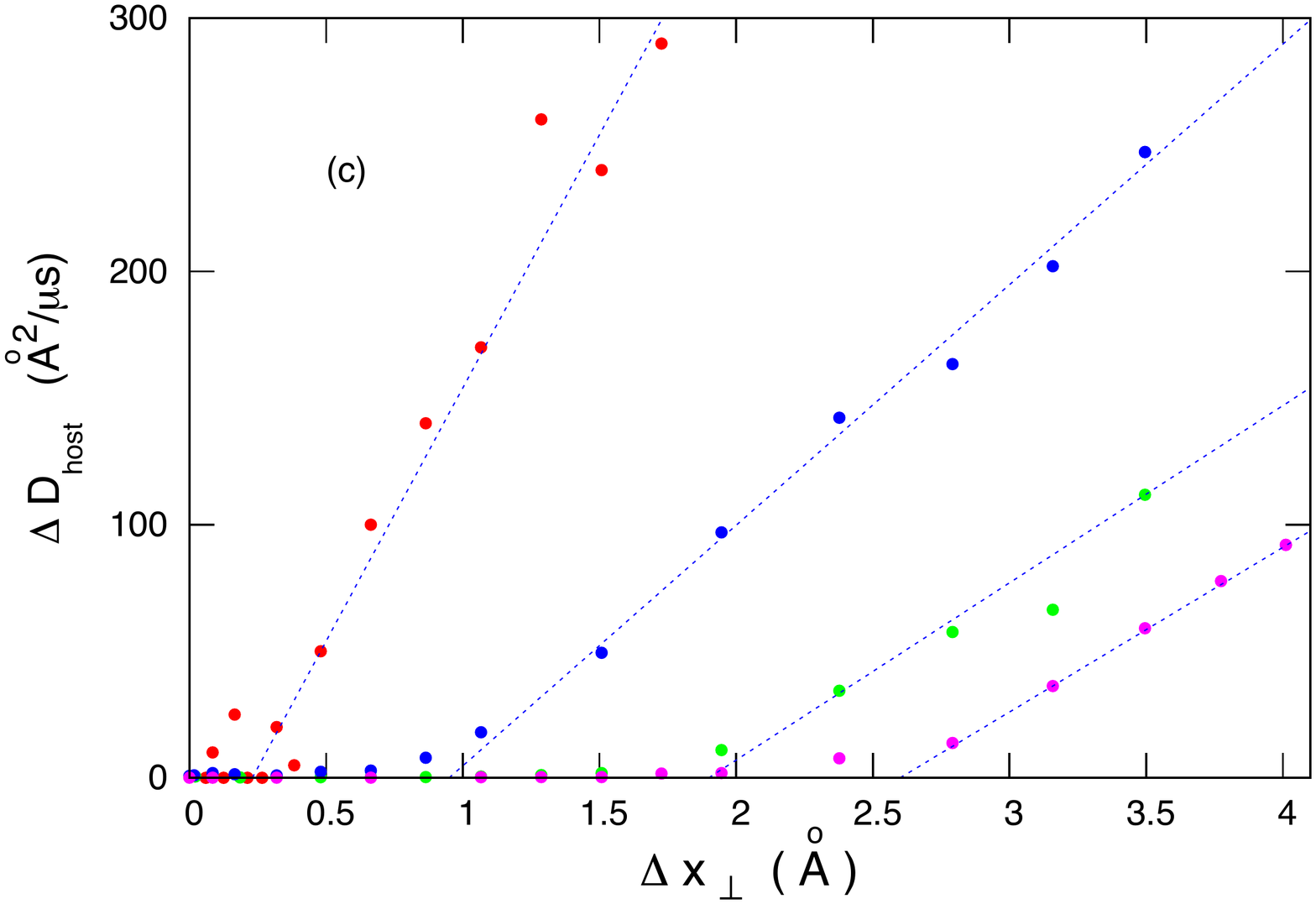}

{\em \footnotesize FIG.3c. (color online) Diffusion coefficient for host molecules, obtained from various isomerization amplitudes $p$ and displayed as a function of $\Delta x_{\perp}$ for different  temperatures. 
From the left to the right hand side: Red full circles: T=40K ($N=300, t _{run} = 200 ns$); Blue full circles: T=30 K   ($N=300, t _{run} = 400 ns$); Green full circles: T=20 K ($N=300, t _{run} = 200 ns$); Purple full circles: T=10 K ($N=300, t _{run} = 200 ns$). For these low temperatures the threshold is more clearly visible.  From this Figure, using linear fits, we find: for T=40K: $\delta x_{\perp}^{threshold}\approx 0.25$\AA, for T=30K: $\delta x_{\perp}^{threshold}\approx 0.95$\AA, for T=20K: $\delta x_{\perp}^{threshold}\approx 1.9$\AA, and for T=10K: $\delta x_{\perp}^{threshold}\approx 2.6$\AA .}

The threshold value increases from $\Delta x_{\perp}=0.25$ \AA\ at T=40K to $\Delta x_{\perp}=2.6$ \AA\ at T=10K. This threshold dependence with the temperature suggests that Brownian thermal motions facilitate the isomerization-induced cage escaping process. In the cage-breaking picture, at the low temperature limit, we expect the threshold to tend to a value  $\Delta x_{\perp}^{limit}\approx2.75$ \AA\ that corresponds to the distance between the first two peaks of the radial distribution function $g_{host-host}(r)$ as discussed above. Consequently the $2.6$ \AA\ value observed at the lowest temperature studied, fits particularly well with the isomerization-induced cage breaking picture.
Above the threshold the diffusion increases linearly with the stimuli $\Delta x_{\perp}$ with a slope that depends on temperature at high enough temperature but then tends to a constant in the low temperatures limit. 
Using a first order Taylor expansion around the threshold we find:

\begin{equation}
D=D_{0}+(\Delta x_{\perp} -\Delta x_{\perp}^{Threshold}) ({ {\partial D} /{\partial \Delta x_{\perp}}})_{{Threshold}}  \label{e1}
\end{equation}

 where $D_{0}$ is the diffusion coefficient at the threshold i.e. the spontaneous diffusion coefficient.
 Then from the relation 
 $D=D_{1}e^{-E_{a}/k_{B}T}$ ($E_{a}(T)$ is the activation energy) that holds for spontaneous diffusion and as $D^{Threshold}=D_{0}$ we obtain:
 
 \begin{equation}
 (\partial D/\partial \Delta x_{\perp})_{{Threshold}}=(-{ {\partial E_{a}}/ {\partial \Delta x_{\perp}}})(D_{0}/k_{B}T)  \label{e2}
 \end{equation}

 The term at the left hand side of this formula is the slope at the very beginning of the lines shown in Figures 3c and 3b. From that formula we obtain a rough estimate of the force $f_{a}$
  that is necessary to induce diffusion (i.e. to break the cage).
 \begin{equation}
 f_{a}=-{\partial E_{a}}/ {\partial \Delta x_{\perp}}  \label{e3}
 \end{equation}
For $T=40K$ in our model we find: $f_{a}/k_{B} \approx 44 K/$\AA.
For lower temperatures the spontaneous diffusion coefficient value  uncertainty is too high to use the formula.
Note that in the low temperature limit the coefficient $D_{0}/k_{B}T$ tends to zero, and as a result the slope around the threshold will tend to zero.
We observe that behavior on the last curve in the right hand side of the Figure 3c, leading to some uncertainty on the position of the threshold for that curve.

Figure 4 shows the mean square displacements (MSD) of the host molecules for various isomerization amplitudes $p$.
The curves display a ballistic behavior at short time scales followed by a plateau characteristic of the molecules caging and finally a diffusive time scale for some curves  while others, at the bottom of the Figure, stay in the plateau regime.  These behaviors are characteristic of supercooled and glassy materials, showing that even when the isomerization is on, the host material located around the chromophore ($R<10$ \AA) behaves as a medium below its melting temperature in the conditions of our simulations. 

\includegraphics[scale=0.33]{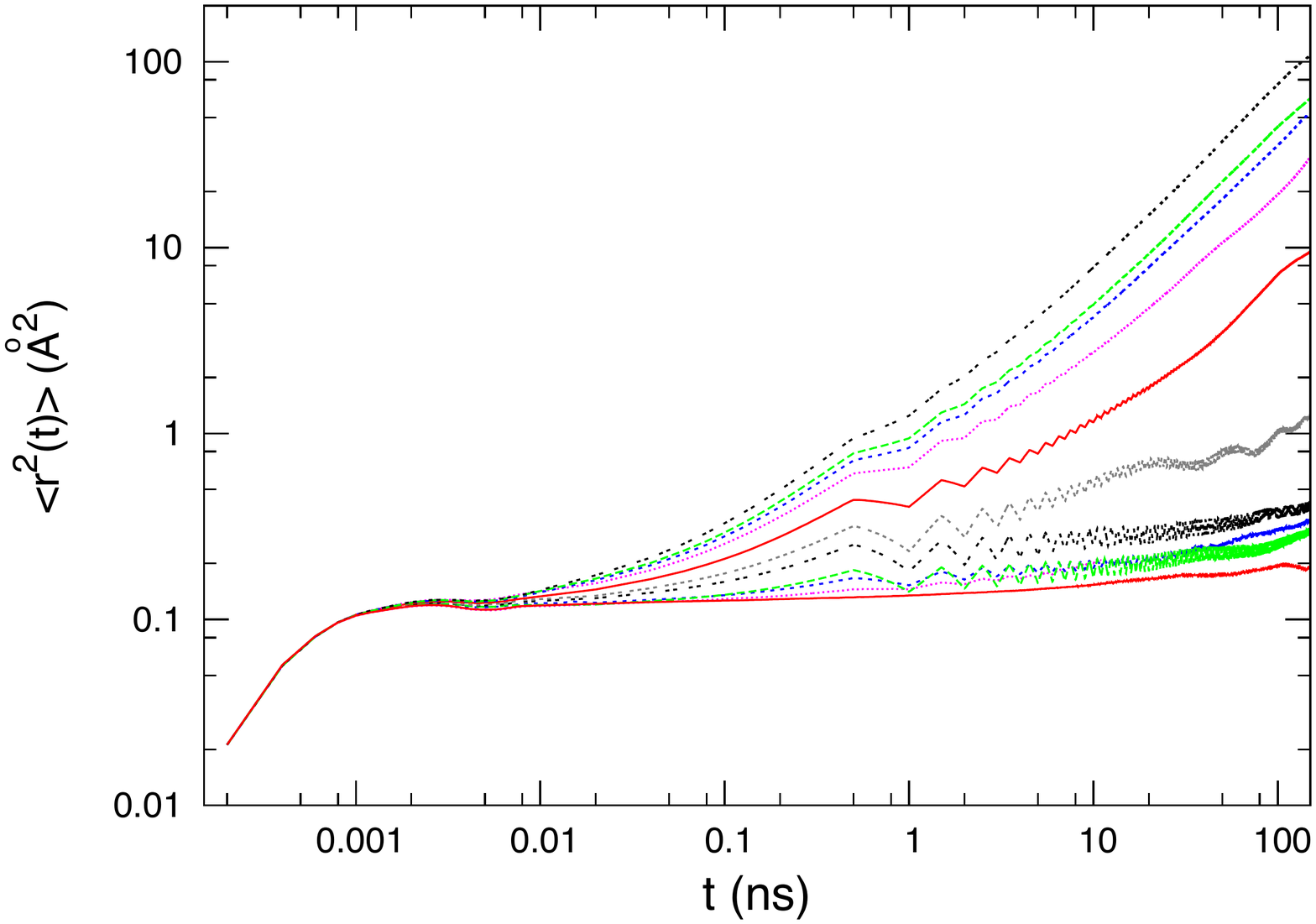}

{\em \footnotesize FIG.4. (color online) Mean square displacement of the host molecules for isomerization amplitudes $p$.  $T=20$ K ($T<T_{g}$) and there is $N=300$ host molecules together with one $DR1 $ molecule in the simulation box.
From bottom to top: Red continuous line: p=$\theta/\theta_{0}=0$; pink dotted line: $p=10\%$; Blue dashed line: $p=20\%$; Green dashed line: $p=30\%$;
Black dashed line: $p=40 \%$; Gray dotted line: $p=50 \%$; Red continuous line: $p=60 \%$; pink dotted line: $p=70\%$; Blue dashed line: $p=80\%$; Green dashed line: $p=90\%$; Black dashed line: $p=100 \%$. The small decrease of the green curve above $50$ ns is a fluctuation. 
Minor tics correspond to a factor 2, 5 and 8 to the major tic they follow.\\}

The first curves  at the bottom of the Figure (corresponding to $p<40\%$) are roughly flat and lead to constant values of the MSD at long time scales.
This behavior shows that the host molecules do not diffuse for $p<40\%$, but we observe oscillations showing that the molecules move periodically around their equilibrium positions.
Figure 4 thus shows that below a threshold $p=40\%$, the isomerizations pushes the host molecules periodically inducing oscillating motions that are not large enough to induce diffusion, i.e.  the molecules move but stay inside their cages.
These oscillations increase with the stimuli $p$ eventually inducing diffusive motions when $\Delta r^{2}>0.4$\AA$^{2}$. 
Interestingly enough that result corresponds to the Lindemann criterion of a vibration amplitude $\Delta r$ reaching approximately $10 \%$ of the nearest neighbor distance $d$ at the melting point as $\Delta r_{threshold}\approx0.2$\AA\  and $d= 2.7$\AA.
Figure 4 also shows that the plateau time range of the mean square displacement decreases when the stimuli increases (i.e. $p$ increases or equivalently $\Delta x_{\perp}$ increases). However the very beginning of the departure from the plateau, i.e. the beginning of the cage escaping process, appears around $10 ps$, while the plateau begins at $t_{0}$=1 ps.
The height of the plateau also increases with the stimuli. This result suggests two possible scenari: i)  The size of the cages (that the plateau height measures) increases with the stimuli. However the constant density of the simulations contradicts that picture. ii)  A few molecules  escape the cages, increasing the mean plateau height, while most molecules still stay inside the cages. In that picture, the number of molecules escaping their cages increases with the stimuli, increasing the mean plateau height.

To gain more insight into the diffusive process, we display in Figures 5 the Self Van Hove correlation functions for the host molecules (Fig.5a and 5c) and for the chromophore (Fig.5b and 5c). 

\includegraphics[scale=0.33]{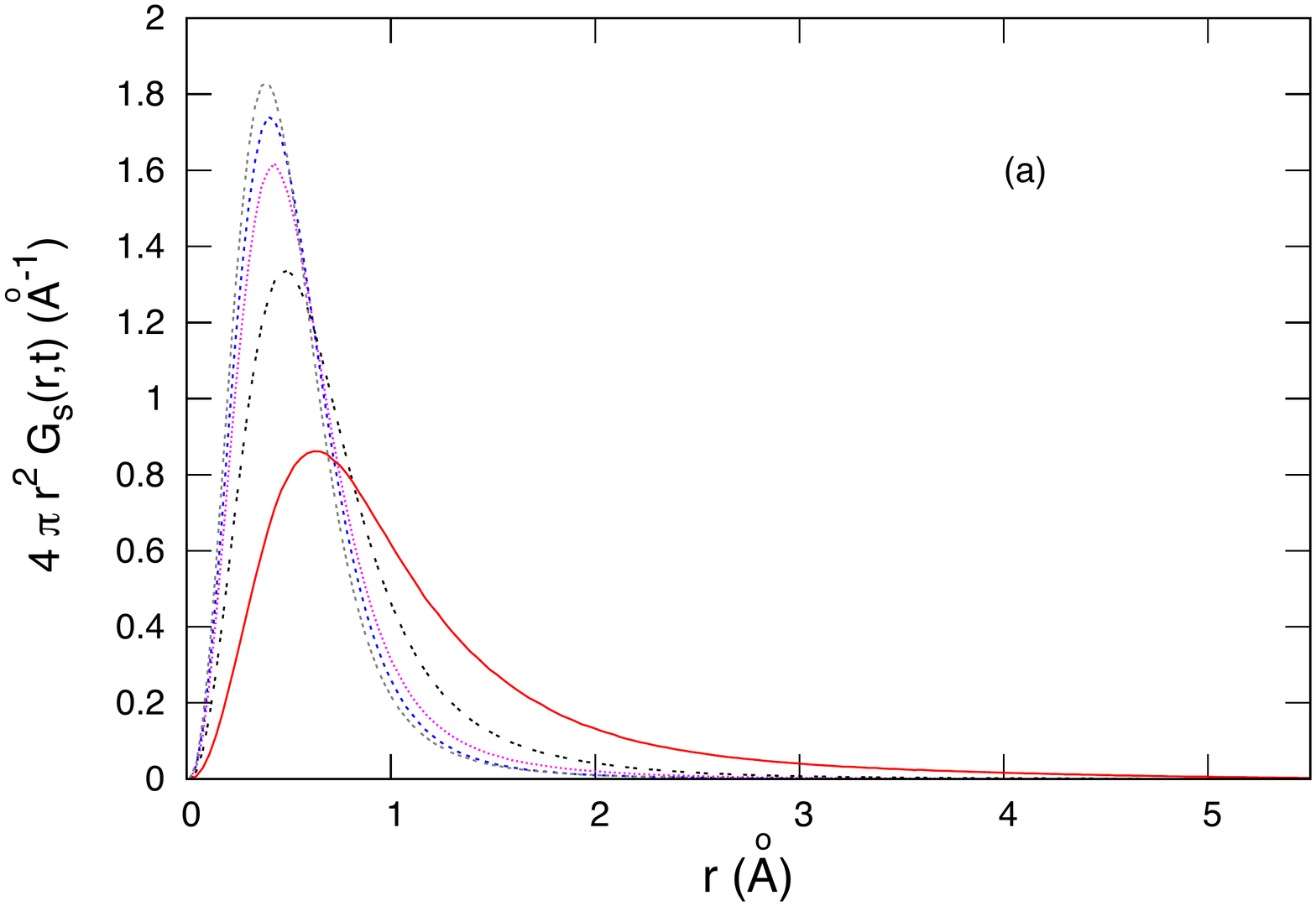}

{\em \footnotesize FIG.5a. (color online) Self Van Hove correlation function $G_{s}(r,t)$ of the host molecules for various isomerization amplitudes $p$. The Van Hove represents the distribution probability to find a molecule at time $t$ (here $t=2.4$ ns) a distance $r$ apart from its position at time zero. T=40 K, N=800 host molecules and one $DR1$.  From the left hand side to the right: Gray dashed curve: $p=0 $; Blue dashed curve: $p=10 \%$; Pink dashed curve: $p=20 \%$; Black dashed curve: $p=30 \%$; Red continuous curve: $p=40 \%$. The molecules do not diffuse below $p=30 \%$.\\}

The Van Hove correlation function measures the distribution probability for a molecule to be a distance $r$ apart from its initial position after a time lapse $t$. Figure 5a shows that the host molecules stay inside their cages for isomerization amplitudes below a threshold value (for $p=\theta /\theta_{0}< 30\% $ the distribution probability stays centered around the same value than when the isomerization is off, and the probability for molecular motions larger than $2.7$\AA\ is almost zero) while they move outside the cages for larger isomerization amplitudes. 

\includegraphics[scale=0.33]{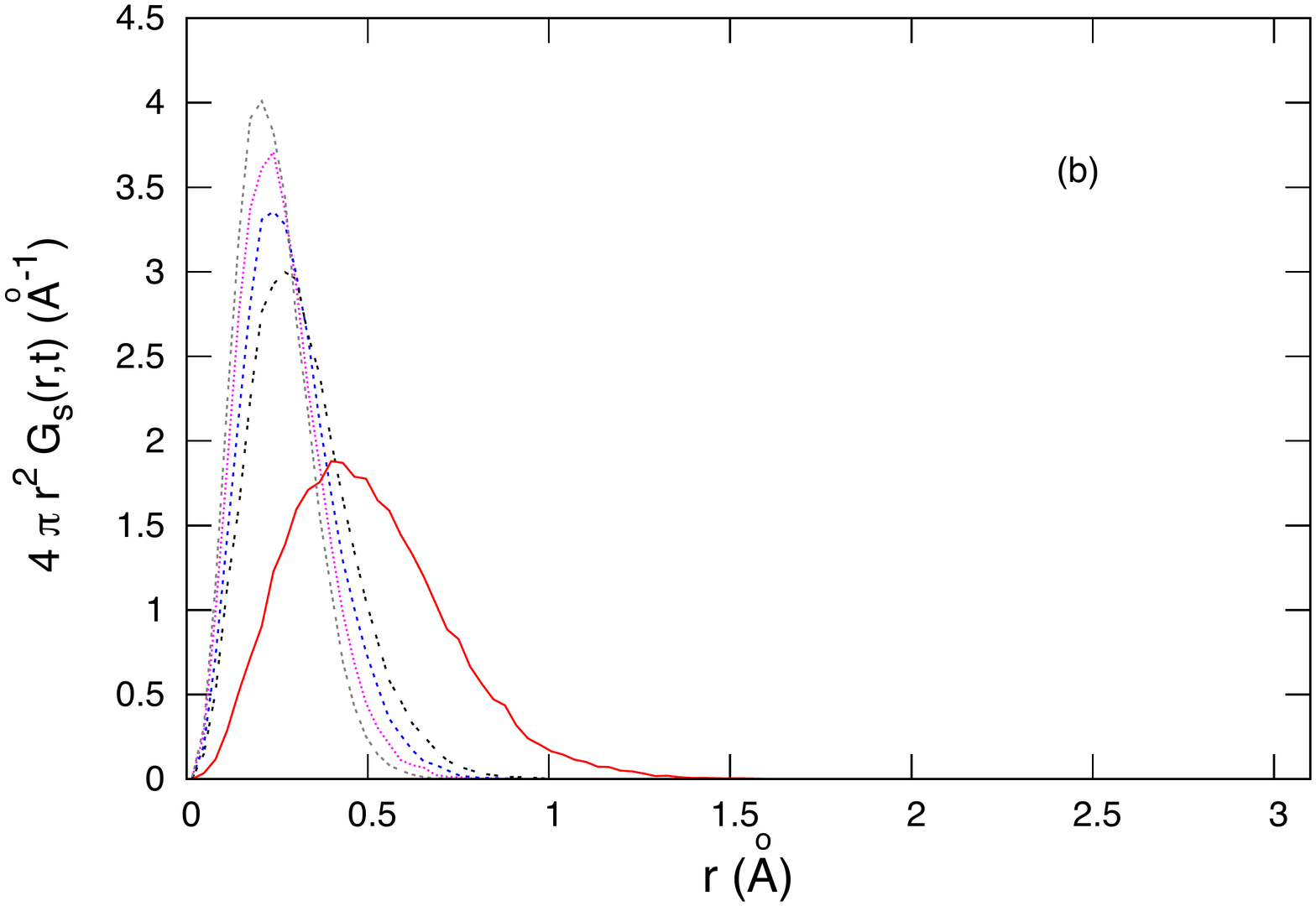}

{\em \footnotesize FIG.5b. (color online) As in Figure 5a but for the chromophore molecule. From the left hand side to the right: Gray dashed curve: $p=0$; Pink dashed curve: $p=10 \%$; Blue dashed curve: $p=20 \%$; Black dashed curve: $p=30 \%$; Red continuous curve: $p=40 \%$. The chromophore does not diffuse below $p=40 \%$.}

The motions of the chromophore (see Figure 5b) display an even more marked isomerization amplitude threshold.
The Figure shows that the chromophore only diffuses above a threshold value $p=\theta/\theta_{0}=40\%$ while the host molecules diffuse for values of $p$ above $30\%$.  We interpret the larger threshold for the chromophore as due to its larger size compared with the host. Due to that larger size of the chromophore more than one host molecules have to move to allow the diffusion of the chromophore outside its cage, leading to a larger threshold for diffusion.  The large mass of the $DR1$ chromophore in comparison with the host molecules also explains the difference in the mobilities of the two sort of molecules when the threshold is passed. 
When compared with the host molecular motions in Figure 5c for a full isomerization ($p=1$; $ \theta =\theta_{0}$) we see that the motion of the chromophore is much smaller than the host motion.  
We also observe the appearance of hopping motions for the chromophore that are not present in the host molecules motions. This behavior is interesting enough as hopping motions are characteristic of solids while continuous diffusive motions are characteristic of liquids. %The chromophore molecule dispersed inside the host liquid thus, due to its large size, interestingly appears to behave as if it was inside a solid host.

\includegraphics[scale=0.33]{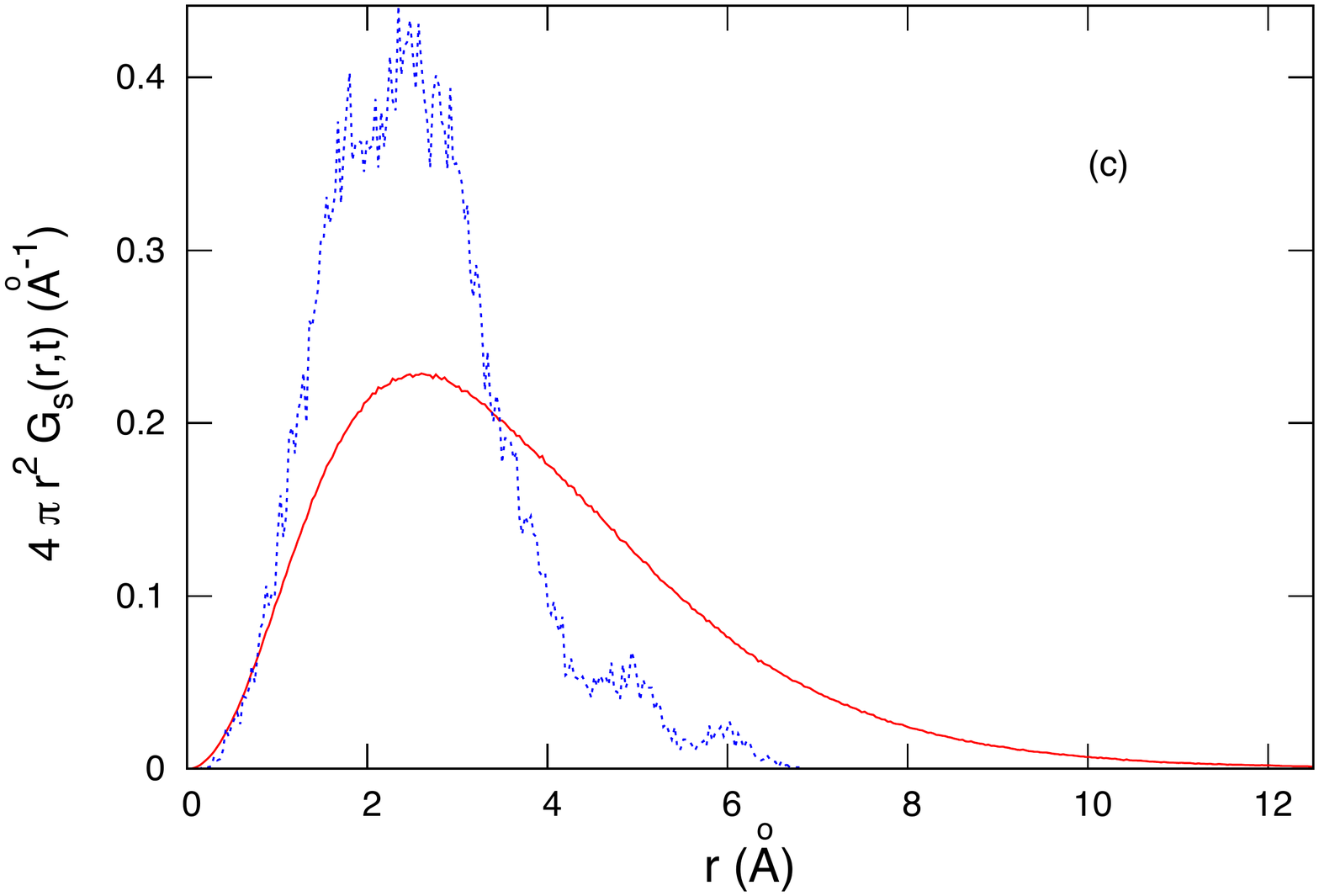}

{\em \footnotesize FIG.5c. (color online) Self Van Hove correlation function $G_{s}(r,t)$ ($t=2.4$ ns) of the host molecules compared to the Van Hove of the chromophore for $p=100\%$.  T=40 K, N=800 host molecules and one $DR1$.   Blue dashed curve: chromophore; Red continuous curve: host. The host  Van Hove function is characteristic of liquids continuous diffusive motions, while hopping solid-like motions are present in the chromophore's Van Hove function.}

\section{Conclusion}

We have studied the effect of a modification of the stimuli on the isomerization-induced transport in azobenzene containing materials. 
We found  that the diffusion does not increase continuously with the stimuli but that there are stimuli thresholds below which the repeated isomerizations do not induce diffusion in the material.
When the temperature decreases we found that the stimuli threshold increases suggesting that thermal activated processes facilitate the diffusion. 
Above the threshold value the diffusion then increases linearly with the stimuli.
The mean square displacements show, below the threshold, induced oscillations of motions that increase with the stimuli.
At the threshold, we found oscillations amplitudes around 13 \% of the nearest neighbor distance ($\Delta r \approx 13\% . d$), a result that reminds the Lindemann criterion for melting.
These results support two models: the induced cage-breaking mechanism\cite{cage} and the gradient pressure theory\cite{a22,a23}, as the origin of the isomerization-induced diffusion in azobenzene containing materials.

\end{document}